\documentclass[a4paper,11pt]{article}
\pdfoutput=1 

\usepackage{jinstpub} 

\title{Characterization of High-Purity Germanium (Ge) Crystals for Developing Novel Ge Detectors}

\author[a ]{M.-S. Raut,}
\author[a] {H. Mei,}
\author[a,1]{D.-M. Mei,\note{Corresponding author.}}
\author[a] {S. Bhattarai,}
\author[a] {W.-Z. Wei,}
\author[a] {R. Panth,}
\author[a] {P. Acharya and}
\author[a] {G.-J. Wang}

\affiliation[a]{Physics Department, University of South Dakota, 414 E. Clark Street, Vermillion, South Dakota, 57069, USA}

\emailAdd{Dongming.Mei@usd.edu}

\abstract{High-purity germanium (HPGe) crystals are required to be well-characterized before being fabricated into Ge detectors. The characterization of HPGe crystals is often performed with the Hall Effect system, which measures the net carrier concentration, the Hall mobility, and the electrical resistivity. The reported values have a strong dependence on the size of the ohmic contacts and the geometry of the samples used in conducting the Hall Effect measurements. We conduct a systematic study using four samples cut from the same location in a HPGe crystal made into different sized ohmic contacts or different geometries to study the variation of the measured parameters from the Hall Effect system. The results are compared to the C-V measurements provided by the Ge detector made from the same crystal. We report the systematic errors involved with the Hall Effect system and find a reliable technique that minimizes the systematic error to be only a few percent from the Hall Effect measurements.        
}

\keywords{Hall Effect, C-V measurement, Carrier Concentration, Rare-event Physics}

\begin{document}
\maketitle
\flushbottom

\section{Introduction}
\label{sec:intro}
In recent years, high-purity germanium (HPGe) crystal detectors have been widely used in searching for dark matter and new properties of neutrinos~\cite{cogent, cdex, cdex1, supercdms, edelweiss, kkdc, gerda, mjd}. These crystals are not only very pure in terms of impurity level down to $\sim$10$^{10}$/cm$^3$, but also extremely pure with respect to the concentration of radioactivity, such as uranium, thorium, and potassium, with an upper limit of $<$10$^{-15}$g/g~\cite{wang, kkdc}. The backgrounds associated with electronics, cryogenic cooling, and shielding have been reduced significantly in last decade. The search for dark mater and new properties of neutrinos utilizing HPGe detectors have entered a new era of a few events with a ton-year of mass exposure. In particular, the search for neutrinoless double-beta decay (0$\upnu$$\upbeta$$\upbeta$) is fundamentally the search for a rare peak superimposed on a background continuum. Therefore, the excellent energy resolution of Ge detectors, far better than noble gas scintillation detectors and probably the best of any 0$\upnu\upbeta\upbeta$ technology, provides highly sensitive discovery potential for the process. This technology also presently has the lowest background when normalized to a resolution-defined region at the $0\upnu\upbeta\upbeta$ Q-value (Q$_{\upbeta\upbeta}$)~\cite{Goeppert}. The detectors are made from pure metallic Ge resulting in a high atomic density and therefore a relatively large number of atoms per kg of detector. Other benefits include the detectors being mostly insensitive to surface activity and the modest cryogenic requirements of liquid nitrogen temperatures. The technology is well established and has been available as a commercial product for many decades.

 To satisfy the high resolution requirements, a HPGe single crystal detector must have a net carrier concentration of $\approx{10^{10}} cm^{-3}$ and a dislocation density between $10^{2}$ etch-pits cm$^{-2}$ and $10^4$ etch-pits cm$^{-2}$ throughout the entire crystal. Production of such crystals is a very challenging task as described by Hansen and Haller~\cite{Hansen}. The zone refinement of commercial Ge ingots to a level of $\approx10^{11} cm^{-3}$ is an essential precursory step to produce detector grade Ge crystals. The University of South Dakota (USD) grows its own HPGe single crystals to meet some stringent criteria for the fabrication of HPGe point-contact detectors. 
 
 One of the several uses of Ge material is the fabrication of HPGe point contact detectors which have a very low energy threshold and an excellent energy resolution over a wide energy range~\cite{Wenzao}. They are in high demand in nuclear and particle physics to probe extremely rare events such as searching for light dark matter and 0$\nu\beta\beta$ decay. Particularly the point contact geometry facilitates pulse-shape discrimination that further reduces backgrounds significantly. The requirement to meet the pulse shape discrimination performance of the point contact detectors is to have an adequate impurity gradient as demonstrated by GERDA and MAJORANA experiments~\cite{gerda, mjd}. Moreover, a point contact detector for light dark matter searches must have a  minimum net carrier concentration as described in a recent paper~\cite{Mei}. This requires that HPGe crystals are precisely characterized in order to understand the electrical properties especially the distribution of carrier concentration.

The carrier concentration, the mobility of the charge carriers, and the resistivity are some of the key electrical properties that characterize  the quality of the HPGe single crystals grown in the laboratory. Electrical characterization of complex samples is increasingly demanded for a vast range of scientific and technical applications.
The van der Pauw Hall effect measurement system is widely used to measure the  above mentioned electrical parameters primarily due to the relative simplicity of the technique and the richness of the information we can obtain. The Hall Effect measurement system is highly encouraged and recommended for new materials for which we do not have much information.
Commonly, a grown crystal is sliced into segments along the growth axis and the samples from certain locations are chosen to conduct the Hall Effect measurements before making a decision on which part of the crystal can be made as a specific type of Ge detector. However, the fabrication of Ge detectors is a meticulous work, requires multiple weeks to process and skilled personnel, stable equipment, and a clean environment~\cite{Wenzao} in order to make a good detector. We often find that if a detector was not made successfully, the origin of the problem can come from the quality of the crystal due to the uncertainty existing in the Hall Effect measurement. Therefore, understanding of the uncertainty in the Hall effect measurement and making improvements to the characterization of grown crystals using the Hall Effect is important. To understand the uncertainty of the Hall Effect measurement, we fabricate grown crystals into small planar detectors. The capacitance versus bias voltage (C-V characteristics) can be used to precisely measure the impurity level. Using the slices cut from the same location of a grown crystal, we can compare the results from the C-V measurement with the Hall Effect system to know the difference and quantitatively characterize the uncertainty of the Hall Effect measurement. For the detector grade crystal grown at USD, we found that the carrier concentration is nearly constant in the central portion of the crystal~\cite{Guojian}. Since the samples we used in this work are all from the central part of the USD-grown crystal, we assume a constant carrier concentration, which is uniformly distributed within the samples used in this work, equals to the net impurity level.   

In this paper, we have cross-checked the reliability and the uncertainty of our Hall Effect measurement by comparing the results to those available from using the C-V method in our laboratory. We have experimented with different van der Pauw geometries to get plenty of data to compare and analyze against C-V results.

\section{Experimental Set Up and Procedures}
At USD, impure Ge metal at a level of $\sim$10$^{14}cm^{-3}$ measured at liquid nitrogen temperature is purified using zone refining techniques. The zone-refined Ge ingots usually reduce the net carrier concentration by 3 orders of magnitude with 14 passes to achieve a level of $\sim$ $10^{11} cm^{-3}$~\cite{Gang1, Gang} at liquid nitrogen temperature. The zone-refined ingots are grown into a single crystal by using the Czochralski method~\cite{Guojian1, Guojian}. The grown crystals are oriented and studied by X-ray diffraction using a Rigaku Ultima IV X-ray diffractometer to understand their orientation. The Olympus BX40 microscope is used to count the dislocation density in the crystals.
The net carrier concentration and material type of selected samples are determined and characterized by using the van der Pauw Hall Effect measurement technique. A very fundamental parameter we measure is the Hall voltage across the sample. Due to a linear dependence on the magnetic field, the carrier concentration ($N$) and the carrier type are derived from the Hall coefficient. The longitudinal resistivity measurement enables us to determine the conductivity and the Hall mobility.

The van der Pauw technique is a non-destructive method which directly measures the sheet resistivity. This method can be easily applied to any arbitrarily shaped samples. Despite all its advantages, it relies on very strict experimental conditions such as infinitely thin samples with point contacts placed on the periphery, and requires homogeneity having no isolated holes within the sample, conditions which are often impossible to attain in practice. This technique is based on the method of conformal mapping and allows for the Hall Effect characterization as well as resistivity measurements in just one experiment with the application of magnetic field~\cite{Pauw}.

In addition to the van der Pauw technique, we use the C-V method to characterize the detector grade sample~\cite{Wei}. By comparing the results obtained from both methods, we can cross-check the reliability and uncertainty in the results obtained from the van der Pauw Hall Effect measurement system.

\begin{figure}[htbp]
\centering 
\includegraphics[width=.5\textwidth,origin=c,angle=0]{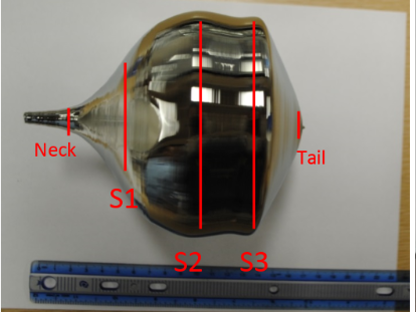}

\caption{\label{fig:i} Shown is a Ge single crystal grown at USD with labelled locations where the samples are usually cut for characterizing the quality of the crystal. S1, S2, and S3 stand for the samples from the shoulder, middle, and bottom of a grown crystal. The region between $S1$ and $S2$ is where the crystal is cut for making a detector. The neck and the tail are the locations of the samples for measurements as well.}
\end{figure}

Although samples of any arbitrary shape and size can be used, for our investigation, square samples of a lateral dimension $l$ and a uniform thickness $t$ are taken from the single Ge detector grade crystal grown at USD, as shown in Figure~\ref{fig:i}. For our specific purpose, we took one set of four samples from upper edge and another set of four samples from lower edge of a single crystal  from where the Ge detector was made. We cut the slices of uniform thickness from the cylindrical crystal using a diamond wire saw.

\begin{figure}[htbp]
\centering 
\includegraphics[width=.8\textwidth,origin=c,angle=0]{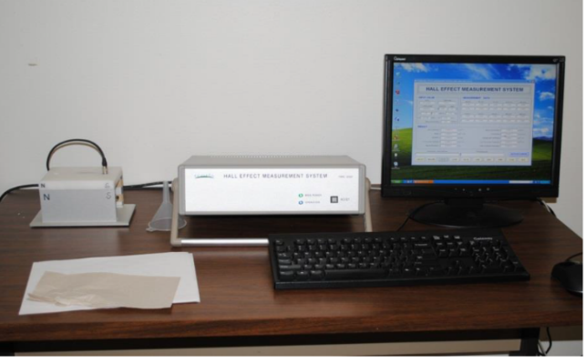}

\caption{\label{fig:ii} Shown is the Ecopia HMS 3000 Hall Effect measurement system.}
\end{figure}

\begin{figure}[htbp]
\centering 
\includegraphics[width=.8\textwidth,origin=c,angle=0]{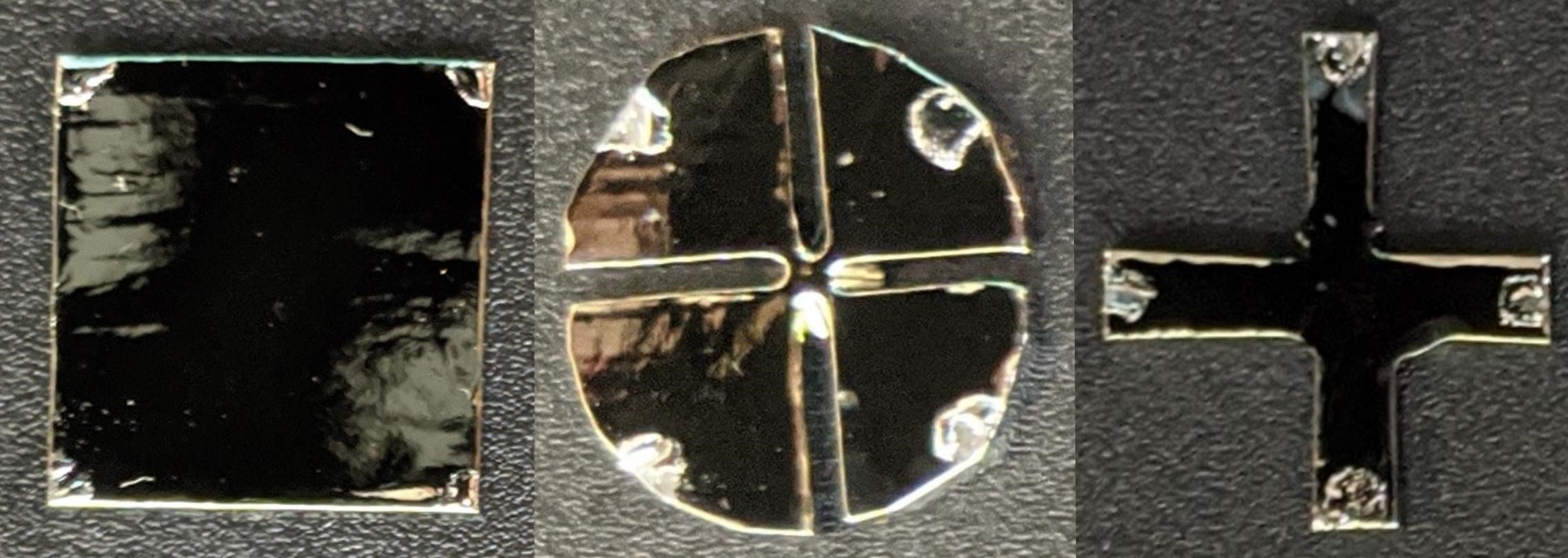}

\caption{\label{fig:iii} Shown is the different forms of the van der Pauw geometry.}
\end{figure}

The obtained slices were again cut into four square symmetric samples from its center region. The samples were etched for three to four minutes thoroughly using an etchant prepared by mixing hydrofluoric acid with nitric acid in the ratio of 1:4 to remove impurities from the surface as well as to make the surface smooth and homogeneous. In the four corners of the samples (p-type), Ohmic contacts were made with the help of Galium Indium eutectic (Ga:In; 75.5:24.5 by weight percentage packed under argon). These van der Pauw samples were heated at 360 degree celsius for half an hour to let Galium Indium eutectic infuse properly into the Ge sample. We measured the samples one by one at liquid nitrogen temperature of 77 K using the Ecopia HMS-3000 Hall Effect equipment equipped with a permanent magnet (0.55 T), as shown in Figure~\ref{fig:ii}. 
After the Hall Effect measurement for the square geometry, the same samples were transformed into Clover leaf and then into Greek cross structures without changing the physical lateral dimensions, as shown in Figure~\ref{fig:iii}.

\subsection{Hall Effect measurement with van der Pauw geometry}
During the measurement, the voltage drop between two contacts is measured while a constant probe current is maintained through the other two contacts. Measuring these resistances on different configurations allows the experimental sheet resistance to be determined using the van der Pauw equation~\cite{Pauw}.

\begin{equation}
\label{eq:1.1}
exp(-\frac{\pi R_1}{R_{s,expt}}) + exp(-\frac{\pi R_2}{R_{s,expt}})  = 1, 
\end{equation}
where $R_{1}$ is the resistance between the first and second contacts when voltage is applied through the third and fourth contacts and $R_{2}$ is that of the third and fourth when voltage is applied between the first and second contacts.
Since we have used symmetric samples with contacts placed on the corners or edges, both resistances will be same and Eq. \eqref{eq:1.1} reduces to a simpler form as,

\begin{equation}
\label{eq:2.2}
R_{s,expt} = \frac{\pi}{ln(2)} {\frac{V}{I}}. 
\end{equation}
The sheet resistance immediately gives the bulk resistivity as, 

\begin{equation}
\label{eq:2.3}
\rho_{s,expt} = R_{s,expt}\times t.
\end{equation}

If a magnetic field $B$ is applied perpendicular to the sample surface, the charge carriers in motion will be deflected by the Lorentz force. The transverse Lorentz force will be compensated by an electric field building up due to the redistribution of mobile charge carriers. In the steady state, the transverse current will be zero, and we measure the Hall voltage between the opposite sides of the sample. This Hall voltage is given by;

\begin{equation}
\label{eq:2.4}
V_H = \frac{R_H B I}{t}. 
\end{equation}
The Hall coefficient is the product of the Hall mobility and the bulk resistivity.  

\begin{equation}
\label{eq:2.5}
R_H = \mu \rho. 
\end{equation}
Once we obtain these parameters, we can calculate the net carrier concentration with the help of following relation. 

\begin{equation}
\label{eq:2.6}
 N = \frac{1}{e R_H}.
\end{equation}
Where $e$ is the electronic charge.

\subsection{C-V measurement}

\begin{figure}[htbp]
\centering 
\includegraphics[width=.6\textwidth,origin=c,angle=0]{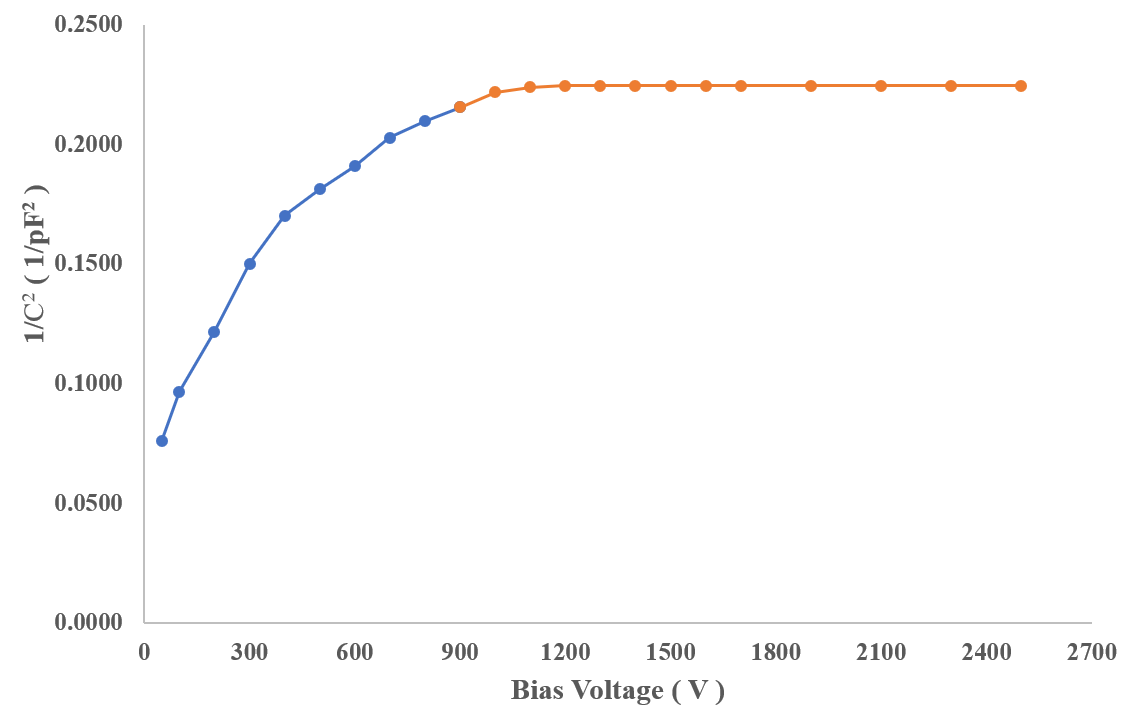}

\caption{\label{fig:iv} Shown is a plot of$ 1/c^2 $vs detector bias (V).}
\end{figure}

The main purpose of the C-V measurements (i.e. detector capacitance as a function of the detector bias voltage) is to determine the full depletion voltage of the detector and in turn the impurity concentration of the crystal~\cite{Wei}. This can be done by applying a small bias voltage step by step from a high voltage (HV) power supply up to a high voltage until the detector is fully depleted at which the capacitance of the detector becomes a constant while continuing to increase the bias voltage. Figure~\ref{fig:iv} shows a C-V curve measurement. 

 The capacitance ($C$) of a planar Ge detector is similar to that of two flat, parallel metallic plates~\cite{Look}. 

\begin{equation}
\label{eq:2.7}
C = \frac{\epsilon_0 \epsilon_{Ge} A}{w}, 
\end{equation}
where $A$ is the active cross-section area of the detector which is $1.40$ $cm^2$, $\epsilon_0 = 8.854\times10^{-14}$ $F/cm$ is the permittivity of free space, $\epsilon_{Ge}$ = 16.2 is the relative permittivity of Ge, and $w$ is the depletion depth of the detector. The depletion depth of the detector is related with applied bias voltage (V) as~\cite{Wei},

\begin{equation}
\label{eq:2.8}
w = (\frac{2\epsilon_0 \epsilon_{Ge} V}{e N(w)})^{1/2}, 
\end{equation}
where $e$ is the elementary charge and $N(w)$ is the net carrier concentration. Note that the net impurity level determined using Eq.~\ref{eq:2.8} may exist some level of uncertainty. A main source of uncertainty is the determination of the depletion voltage using the C-V method. As shown in Figure~\ref{fig:iv}, when the detector is fully depleted, the capacitance is a constant. The depletion voltage is thus chosen using the C-V curve where the bias voltage corresponds to the first point of a constant capacitance. This uncertainty was evaluated to be less than 1\% when the bias voltage is slowly increased at a rate of every 10 volts for a planar geometry.  

For the detector grade crystal, the impurity concentration is assumed to be constant and the doping to be uniform~\cite{Guojian}. With this assumption, as the applied voltage (V) increases, the depletion depth ($w$) increases and the capacitance (C) decreases until we achieve full depletion. Once the full depletion occurs, the capacitance (C) and the depletion depth (w) remain constants on further increment of applied bias voltage (V). At this moment the constant depletion depth becomes the detector thickness and the net carrier concentration(N(n or p)) will be equal to the impurity/dopant concentration (N(D) or N(A)). Hence, using eq.\eqref{eq:2.8}, we can calculate the net carrier concentration N(w) as we obtain the bias voltage (V) at full depletion.
For our case, the detector thickness (w) is 0.94 cm and depletion voltage (V) is 1300 V. Thus, the net carrier concentration (N(w)) is 2.639$\times$10$^{10}$ cm$^{-3}$.

\section{Results and Discussion}
The raw data measured for the four sides of a square sample (the lateral dimension ($l$) and the diameter for the contact size ($c$ )) of the four samples are given in Table~\ref{tab:i}. The mean in the second last row of the table is the average of the respective data set for the four samples. The subscript to the mean is the value obtained by subtracting the same mean value from the minimum entry of the data set. Similarly, the superscript is the value obtained by subtracting the mean from the maximum entry of the data set. The last row of the table shows the aspect ratio of the respective samples. The errors assigned to the aspect ratio in subscript and in superscript are obtained by taking the largest deviations from either side of the mean value. The errors calculated for the aspect ratio ($c/l$) are the maximum possible errors on either side of the mean value associated with each of the contact diameter ($c$) and the lateral dimension ($l$).
\begin{table}[htbp]
\centering
\small\addtolength{\tabcolsep}{-5pt}
\caption{\label{tab:i} The aspect ratio of the samples with smaller contacts.}
\smallskip
\begin{tabular}{|c|c|c|c|c|c|c|c|c|}
\hline
Sample &  \multicolumn{2}{|c|}{S1-1} & \multicolumn{2}{|c|}{S1-2} & \multicolumn{2}{|c|}{S1-3} & \multicolumn{2}{|c|}{S1-4}\\
\hline
&c (mm)&l (mm)& c (mm)&l (mm) &c (mm)&l (mm)&c (mm)&l (mm)\\
\hline
&2.95&16.68&	2.64&	16.33&	3.45&	16.33&	3.22&
14.44\\
&3.45&	17.47&	2.55&	16.35&	2.78&	16.81&	2.84&	15.76\\
&2.81&	16.57&	3.49&	16.43&	2.79&	16.28&	2.92&	13.44\\
&3.00&	16.53&	3.12&	16.78&	2.91&	16.71&	3.07&	16.03\\
\hline
Mean & $3.05_{-0.24}^{+0.40}$&	$16.81_{-0.28}^{+0.66}$&	$2.95_{-0.40}^{+0.54}$&	$16.47_{-0.14}^{+0.31}$&	$2.98_{-0.20}^{+0.47}$&	$16.53_{-0.25}^{+0.28}$&	$3.01_{-0.17}^{+0.21}$&	$14.92_{-1.48}^{+1.11}$\\
\hline
c/l &  \multicolumn{2}{|c|}{$0.18_{-0.02}^{+0.02}$} & \multicolumn{2}{|c|}{$0.18_{-0.03}^{+0.03}$} & \multicolumn{2}{|c|}{$0.18_{-0.01}^{+0.03}$} & \multicolumn{2}{|c|}{$0.20_{-0.02}^{+0.04}$}\\

\hline
\end{tabular}
\end{table}\\

Using the Ecopia Hall Effect system, we measured the bulk resistivity and mobility for the samples. The obtained results for the four samples from wafer slice S1 are tabulated in Table~\ref{tab:ii} for illustration. Table~\ref{tab:iii} enlists the carrier concentrations for different van der Pauw geometries for two different aspect ratios. The mean in the last row of both the tables is the average of the respective data set for four samples. The subscript to the mean is the value obtained by subtracting the same mean value from the minimum entry of data set. Similarly, the superscript is the value obtained by subtracting the mean from the maximum entry of the data set.

\begin{table}[htbp]
\centering
\small\addtolength{\tabcolsep}{-5pt}
\caption{\label{tab:ii} The resistivity and mobility of the square samples with smaller contacts. (Here, the resistivities are in $\times$10$^3$  $\Omega$ cm and the mobilities are in $\times$ 10$^4$ cm$^2$/$Vs$.}
\smallskip
\begin{tabular}{|c|c|c|c|c|c|c|c|c|}
\hline
Sample &  \multicolumn{2}{|c|}{S1-1} & \multicolumn{2}{|c|}{S1-2} & \multicolumn{2}{|c|}{S1-3} & \multicolumn{2}{|c|}{S1-4}\\
\hline
Parameter& Resistivity&Mobility& Resistivity&Mobility&Resistivity&Mobility&Resistivity&Mobility\\
\hline

&6.24&	4.95&	5.56&	4.86&	1.46&	4.29&	8.37&	4.03\\
&5.90&	5.46&	6.58&	4.57&	6.87&	4.27&	3.95&	3.96\\
&9.23&	4.38&	6.23&	3.99&	5.59&	4.89&	6.66&	4.67\\
&5.35	&3.97&	6.00&	5.21&	5.90&	3.87&	6.78&	5.36\\
&5.67&	4.67&	4.65&	4.63&	6.73&	5.00&	8.37&	4.59\\
\hline
Mean& $6.45_{-1.13}^{+2.75}$&	$4.69_{-0.72}^{+0.77}$&	$5.81_{-1.15}^{+1.65}$&	$4.65_{-0.67}^{+0.56}$&	$5.31_{-3.85}^{+3.79}$&	$4.47_{-0.59}^{+0.54}$&	$6.83_{-2.87}^{+1.54}$&	$4.52_{-0.56}^{+0.83}$ \\
\hline
\end{tabular}
\end{table}

\begin{table}[htbp]
\centering
\small\addtolength{\tabcolsep}{-5pt}
\caption{\label{tab:iii} The mean carrier concentration (in $\times$10$^{10}$ per cm$^3$) of the samples for three different van der Pauw geometries with smaller and larger contacts on them as measured by our Hall Effect measurement system.}
\smallskip
\begin{tabular}{|c|c|c|c|c|c|c|}
\hline
Sample &  \multicolumn{2}{|c|}{Square} & \multicolumn{2}{|c|}{Greek} & \multicolumn{2}{|c|}{Clover leaf} \\
\hline
&Larger&Smaller& Larger&Smaller&Larger&Smaller\\
\hline
S1-1&	7.23&	2.06&	3.78&	3.55&	4.46&	2.31\\
S1-2&	1.99&	2.31&	1.97&	2.67&	2.88&	2.75\\
S1-3&	2.66&	2.63&	8.31&	3.00&	1.69&	2.20\\
S1-4&	2.75&	2.02&	2.09&	2.71&	2.15&	3.09\\
\hline
Mean&	$3.66_{-1.67}^{+3.57}$&	$2.26_{-0.23}^{+0.38}$&	$4.04_{-2.07}^{+4.27}$&	$2.98_{-0.31}^{+0.57}$&	$2.79_{-1.10}^{+1.67}$&	$2.59_{-0.39}^{+0.50}$\\
\hline
\end{tabular}
\end{table}

We repeated the same process for wafer S2 and calculated results for the net carrier concentration are averaged with the wafer S1 for each respective van der Pauw structure to obtain mean net carrier concentration.
The mean net carrier concentrations as obtained from the Hall Effect measurement with different aspect ratios for different van der Pauw geometries are listed in Table~\ref{tab:iv}.
\begin{table}[htbp]
\centering
\small\addtolength{\tabcolsep}{-5pt}
\caption{\label{tab:iv} The aspect ratio and the carrier concentration for smaller and larger contacts in three different van der Pauw geometries.}
\smallskip
\begin{tabular}{|c|c|c|c|c|}
\hline
Geometry&Contact Size & c/l & Concentration(N) per $cm^3$ \\
\hline
Square & Larger &0.23&	3.79$\times10^{10}$ \\
&Smaller&	0.19&	2.06$\times 10^{10}$\\
\hline
Greek Cross&Larger&	0.26&	4.20$\times 10^{10}$\\
&Smaller	&0.18&3.01$\times 10^{10}$\\
\hline
Clover Leaf&Larger&	0.25&	2.99$\times 10^{10}$\\
&Smaller	&0.20&2.63$\times 10^{10}$\\
\hline
\end{tabular}
\end{table}
From Table~\ref{tab:iii}, it is clearly seen that the observed values for concentration for larger contacts are consistently higher than those observed for smaller contacts in each kind of van der Pauw geometry.

We compared these results with the result obtained from our C-V measurement (Table~\ref{tab:v}). We found that a higher aspect ratio of the samples will result in a larger percent difference and vice-versa. Furthermore, we can see the percent difference for the clover leaf structures is the smallest when compared to those of Greek Cross and regular square structures.

\section{Correction and Error Analysis}
\subsection{The van der Pauw Correction}
The experimental sheet resistance is equal to the true sample sheet resistance if all of the ideal conditions and assumptions are fully satisfied. In reality, one or more non-ideal conditions are usually present. In such cases, the experimental sheet resistance differs from the true sample resistance by a factor,  which depends on the specific sample geometry, that needs to be corrected. We can define this correction factor as~\cite{Pauw},

\begin{equation}
\label{eq:4.1}
f(Corr.)=\frac{R_s}{R_{s,expt}}.
\end{equation}
\begin{table}[htbp]
\centering
\caption{\label{tab:v} A summary of the errors associated with different samples. }
\smallskip
\begin{tabular}{|c|c|c|c|c|c|c|}
\hline
Geometry &  \multicolumn{2}{|c|}{Square} & \multicolumn{2}{|c|}{Greek Cross} & \multicolumn{2}{|c|}{Clover leaf}\\ 
\hline
Contact Size&Larger&Smaller& Larger&Smaller&Larger&Smaller\\
\hline
Percent difference&	35.91&	24.83&	45.76&	13.19&	12.57&	0.76\\
\hline
\end{tabular}
\end{table}
Most of the intrinsic errors in the Hall Effect measurement system are compensated on by the magnetic field reversal and the probe current reversal that we apply during the measurement. However, we cannot ignore the effects of contact size and hence the aspect ratio ($c/l$) in the calculation of the bulk resistivity and the Hall coefficient. In this work,  we used the van der Pauw correction relations for them respectively. 
\subsubsection{Correction for square geometry}
The van der Pauw correction in the bulk resistivity is given by:

\begin{equation}
\label{eq:4.2}
\frac{\Delta \rho}{\rho}=-\frac{ln(1+\frac{(c/l)^2}{(1-(c/l)^2})}{2 ln 2}
\end{equation} ~\cite{Pauw}.
So, this is approximated as:

\begin{equation}
\label{eq:4.3}
\frac{\Delta \rho}{\rho}\approx(c/l)^2. 
\end{equation}
Eq.~\ref{eq:4.3} indicates that the ratio of the increase of the resistivity to the resistivity is proportional to the squared aspect ratio.  
Similarly, the Hall coefficient is corrected as~\cite{Pauw}:

\begin{equation}
\label{eq:4.4}
\frac{\Delta R_H}{R_H}\approx(c/l).  
\end{equation}

The Hall coefficient decreases with increase in aspect ratio. 
When we apply these corrections for the Hall Effect results in Table~\ref{tab:iv}, the percent difference we obtained are tabulated below in Table~\ref{tab:vi}.
\subsubsection{Correction for Greek-cross geometry}
This geometry is one of the best van der Pauw geometries to minimize finite contact errors.The deviation of the actual resistivity $\rho$ has been found to be deviated from the measured one given by the following numerical relation~\cite{David}.
\begin{equation}
\label{eq:4.5}
E = 1-\frac{\rho_0}{\rho_m}= (0.59\pm 0.006) exp[-(6.23\pm 0.02) \frac{a}{c}]  
\end{equation}
Where $a$ is the arm length of Greek-cross structure and $c$ is contact dimension which are related with lateral dimension of sample by $c+2a$ = $l$. In our study, the typical value for aspect ratio is around $0.17$ which produces very small error $E$ $\approx{10^{-7}}$.
For the error associated with the Hall Coefficient, De Mey has devised the following relation for four contacts~\cite{DeMey}.
\begin{equation}
\label{eq:4.6}
\frac{\mu_0 - \mu_m}{\mu_0} \approx{1.045 exp[-\pi a/c]}  
\end{equation}
Where $\mu_0$ and $\mu_m$ are the actual and measured Hall mobilities respectively. With our typical value of aspect ratio, this error is approximately $0.038\%$.
\subsubsection{Correction for Clover leaf geometry}
Clover leaf geometry is circular van der Pauw structure. The correction factor provided by van der Pauw per contact for this structure is given~\cite{Pauw}:
\begin{equation}
\label{eq:4.7}
\frac{\Delta \rho}{\rho} \approx {-\frac{1}{16 ln2}(\frac{c}{l})^2}
\end{equation}

Which results around $-1.09\%$ error in resistivity $\rho$ for our typical aspect ratio.
Similarly, for the Hall coefficient, van der Pauw provides the correction per contact~\cite{Pauw} as:
\begin{equation}
\label{eq:4.8}
\frac{\Delta R_H}{R_H} \approx {\frac{2}{\pi ^2} \frac{c}{l}}
\end{equation}
For aspect ratio of $0.17$, this results in a correction of $11.57\%$ for four contacts.

With all these corrections applied to the respective structures, the obtained percent differences are tabulated in table~\ref{tab:vi}.

\subsection{Comparison}
When we compare the results obtained from samples with a higher aspect ratio to that of having a smaller aspect ratio, the reliability of the results is higher in latter with a smaller uncertainty.
\begin{table}[htbp]
\centering
\caption{\label{tab:vi} A comparison of the errors between the square samples.  }
\smallskip
\begin{tabular}{|c|c|c|c|c|c|c|}
\hline
Geometry &  \multicolumn{2}{|c|}{Square} & \multicolumn{2}{|c|}{Greek Cross} & \multicolumn{2}{|c|}{Clover leaf}\\ 
\hline
Aspect Ratio&Larger&Smaller& Larger&Smaller&Larger&Smaller\\
\hline
Percent difference&	14.55&	3.04&	23.71&	9.88&	7.24&	1.37\\
\hline
\end{tabular}
\end{table}

\section{Conclusion}
From this study, we found that the Hall Effect measurement is largely affected by the aspect ratio and the van der Pauw geometry of the samples. Clover leaf structures, even without correction, give a closer value to the C-V result. Square samples with a smaller aspect ratio yield acceptable and reasonable error. The human error in measuring the length of the samples, the systematic error in the calculations of these values assuming that the samples are homogeneous without holes to meet the van der Pauw experimental conditions,  and the random error in the equipment calibration justify the percent difference observed in Table~\ref{tab:vi}.
We also have assumed that there are no offset voltages and offset currents arising due to an improperly zeroed voltmeter and ammeter in our Hall Effect measurement system. A temperature gradient across the sample allows two contacts to function as a pair of thermo-couple junctions. The resulting thermoelectric voltage due to the Seebeck effect is not affected by current or magnetic field, to first order. Even if no external transverse temperature gradient exists, the sample can set up its own. The Lorentz force shunts slow (cold) and fast (hot) electrons to the sides in different numbers and causes an internally generated Seebeck effect. This phenomenon is known as the Ettingshausen effect~\cite{Look}. Unlike the Seebeck effect, it is proportional to both current and magnetic field.

If a longitudinal temperature gradient exists across the sample, the electrons tend to diffuse from the hot end to cold end of the sample and this diffusion current is affected by a magnetic field, producing a Hall voltage. The phenomenon is known as the Nernst effect. Thus resulting voltage is proportional to magnetic field but not to external current. This is the one source of intrinsic error which cannot be eliminated by magnetic field or current reversal. Even in zero magnetic field, a voltage appears between the two contacts used to measure the Hall voltage if they are not electrically opposite each other. Voltage contacts are difficult to align exactly. The misalignment voltage is frequently the largest spurious contribution to the apparent Hall voltage.

Though the clover leaf structures seem best to use, these structures are very fragile to handle and tedious to make. We conclude that square samples with smaller contacts after using the van der Pauw correction to the Hall coefficient and the bulk resistivity can give reasonable results with small errors for the Hall Effect measurement system.  

\section*{Acknowledgments}
The authors would like to thank Mark Amman for his instructions on fabricating planar detectors and Christina Keller for a careful reading of this manuscript. We would also like to thank the Nuclear Science Division at Lawrence Berkeley National Laboratory for providing us a testing cryostat. This work was supported in part by NSF NSF OISE 1743790, NSF PHYS 1902577, NSF OIA 1738695, DOE grant DE-FG02-10ER46709, DE-SC0004768, the Office of Research at the University of South Dakota and a research center supported by the State of South Dakota.

\end{document}